\documentclass[aps,prd,amsmath,twocolumn,amssymbaps,showpacs]{revtex4-1}
\usepackage{graphicx}  
\usepackage{dcolumn}   
\usepackage{bm}        
\usepackage{amssymb}   
\usepackage{amsmath}   
\usepackage{lineno}
\usepackage{draftcopy}
\usepackage{relsize} 
\usepackage{xfrac}    
\usepackage{slashed}

\usepackage{color}
\definecolor{dgreen}{cmyk}{1.,0.,1.,0.2}        
\definecolor{orange}{cmyk}{0.,0.353,1.,0.}    

\usepackage[bookmarks]{hyperref}

\newcommand{\be}{\begin{equation}}
\newcommand{\ee}{\end{equation}}                                                                               
\newcommand{\bea}{\begin{eqnarray}}
\newcommand{\eea}{\end{eqnarray}}

\begin{document}
\title{Effects of a parallel electromagnetic field in three-flavor Nambu--Jona-Lasinio model}

\author{Gaoqing Cao}
\affiliation{School of Physics and Astronomy, Sun Yat-sen University, Guangzhou 510275, China.} 
\date{\today}

\begin{abstract}
In this work, we explore the ground state of QCD system and the relevant collective modes in a parallel electromagnetic (EM) field within the effective three-flavor Nambu--Jona-Lasinio model. From the features of neutral chiral condensates, three critical EM fields are identified in our study: $e\bar{I}_2^{c_i}~(i=1,2,3)$. Moreover, competition between QCD and QED anomalies is found: the negativeness of $\pi_{\rm d}^*$ and $\pi_{\rm s}^*$ is a signal of QCD anomaly dominance for all the flavors, and the positiveness of $\pi_{\rm s}^*$ beyond $e\bar{I}_2^{c_1}$ indicates QED anomaly dominance for strange quark. For the lowest-lying neutral collective modes, the masses of $\Pi$ and $H$ modes reduce to zero around the critical EM fields and the scalar-pseudoscalar components are exchanged between $\Sigma$ and $H$ modes, as can be seen in the crossing structure.
\end{abstract}


\maketitle

\section{Introduction}
In recent years, a lot of scientific interest has focused on the properties of QCD systems under extreme conditions, including those existing in certain types of astrophysical objects and those created in relativistic heavy-ion collisions. Large baryon  density (of the same order as the saturation density $\rho_0=0.16~{\rm fm}^{-3}$) is expected to occur in neutron stars, especially the inner cores. Pasta structures~\cite{Ravenhall:1983uh,Maruyama:2006jk}, chiral symmetry restoration~\cite{Klevansky:1992qe,Hatsuda:1994pi}, partial-wave nucleon pairings~\cite{Sedrakian:2018ydt,Fujimoto:2019sxg} and color superconductivity~\cite{Alford:1997zt,Rapp:1997zu} have been extensively studied in such circumstance. These explorations focus mainly on the equilibrium ground states of the QCD system which are related to the equation of state (EoS). Actually, there is some ambiguity in the determination of the EoS but some new insights have been gained from recent multi-messenger observations of binary neutron star mergers: GW170817~\cite{GBM:2017lvd}. As a matter of fact, detection of the associated gravitational wave (GW) constrains the tidal deformabilities and the radii of neutron stars to $\tilde{\Lambda}<800$ and $R=8.9-13.2{\rm km}$~\cite{De:2018uhw,Most:2018hfd}, and the combined optical and GW observations of GW170817 constrain the maximum mass of neutron stars to $2.16-2.28M_\odot$~\cite{Margalit:2017dij,Ruiz:2017due} and the EoS is found to be slightly soft at large baryon density $\rho>5\rho_0$~\cite{Radice:2017lry,Abbott:2018exr}. 

In the large terrestrial facilities, very strong electromagnetic (EM) field ($10^{18}-10^{20}G$)~~\cite{Skokov:2009qp,Deng:2012pc,Deng:2014uja,Bloczynski:2012en} and fast rotation ($\sim 10^{22} s^{-1}$)~\cite{STAR:2017ckg} can be produced in peripheral heavy ion collisions (HICs), and the chiral magnetic effect (CME)~\cite{Liao:2014ava,Kharzeev:2015znc,Huang:2015oca,Bu:2019qmd} and particle polarization effect~\cite{Liang:2004ph,Karpenko:2016jyx,Xia:2018tes,Wei:2018zfb,Guo:2019joy,Becattini:2019ntv,Xia:2019fjf,Wu:2019eyi,Xia:2019whr,Niida:2018hfw,Adam:2019srw} are hot topics on both theoretical and experimental aspects. Besides, the effects of an EM field and rotation on QCD ground states and the associated extended phase diagrams have been widely studied and several intriguing phenomena were discovered, such as "inverse magnetic catalysis effect"~\cite{Bali:2011qj,Bali:2012zg}, "anomalous neutral pion superfluidity"~\cite{Cao:2015cka,Wang:2018gmj,Wang:2017pje,Cao:2019hku}, "magnetic anisotropic confinement"~\cite{Bonati:2014ksa,Bonati:2016kxj,Cao:2019azh}, and "rotational magnetic inhibition"~\cite{Chen:2015hfc}. Directly related to the ground states, neutral and charged meson masses have also been studied in different scenarios~\cite{Hidaka:2012mz,Bali:2017ian,Ding:2020jui,Avancini:2016fgq,Wang:2017vtn,Mao:2018dqe,Liu:2018zag,Coppola:2018vkw,Cao:2019res}. Take two important discoveries for example: the charged rho meson mass never reduces to zero in pure magnetic field according to the lattice QCD simulations~\cite{Hidaka:2012mz,Bali:2017ian,Ding:2020jui}, but the charged pion meson mass might vanish under the condition of parallel magnetic field and rotation~\cite{Liu:2017spl,Cao:2019ctl,Chen:2019tcp}.

Among all the circumstances explored, the case with parallel EM (PEM) field is special because neutral pseudoscalar condensations can be induced through chiral anomaly~\cite{Cao:2015cka,Wang:2018gmj,Wang:2017pje,Cao:2019hku}. On the experimental side, PEM field was used to produce chirality imbalance in the Weyl semimetal system and the well-known CME was discovered for the first time~\cite{Li:2014bha}. Besides, it was found that the observed charge-dependent elliptic flow of pions might be understood by the PEM field distribution in HICs~\cite{Zhao:2019ybo}. Present work is just the three-flavor version of our previous researches~\cite{Cao:2015cka,Wang:2018gmj,Wang:2017pje,Cao:2019hku} -- the advantages are that this case is more realistic and the QCD $U_A(1)$ anomaly is automatically accounted for through the 't Hooft determinant~\cite{Klevansky:1992qe,Hatsuda:1994pi}. With the latter point, the relative significances and effects of QCD and QED anomalies can be well compared in the three-flavor case. As PEM field is indeed relevant in HICs~\cite{Zhao:2019ybo}, we hope that in the future more attention will be paid to the possibility
of anomalous pseudoscalar condensations in experiments besides the CME signals. 

The paper is arranged as follows. In Sec.~\ref{formalism}, we present the whole formalism for the studies of chiral condensates and collective modes, with gap equations in Sec.~\ref{gapequation} and polarization functions in Sec.~\ref{Pfunction}. The numerical results are shown in Sec.~\ref{num} and finally we conclude in Sec.~\ref{conclusion}.

\section{Three-flavor formalism}\label{formalism}
\subsection{Gap equations}\label{gapequation}
In order to improve upon previous studies~\cite{Cao:2015cka,Wang:2018gmj,Wang:2017pje,Cao:2019hku} and study QCD system more
realistically, we adopt the three-flavor Nambu--Jona-Lasinio (NJL) model where more low-lying collective modes are involved and the QCD $U_A(1)$ anomaly has been taken into account automatically through the 't Hooft term. The corresponding Lagrangian density is~\cite{Klevansky:1992qe,Hatsuda:1994pi}:
\begin{eqnarray}
{\cal L}_{\rm NJL}&=&\bar\psi(i\slashed{D}-m_0)\psi+G\sum_{a=0}^8\left[(\bar\psi\lambda^a\psi)^2+(\bar\psi i\gamma_5\lambda^a\psi)^2\right]\nonumber\\
&&+{\cal L}_{\rm tH},\  \ {\cal L}_{\rm tH}=-K\sum_{s=\pm}{\det}~\bar\psi\Gamma^s\psi,\label{LNJL}
\end{eqnarray}
where $\psi=(u,d,s)^T$ represents the three-flavor quark field and $ {\cal L}_{\rm tH}$ is the 't Hooft term with coupling $K$. In the kinetic term, $m_0={\rm diag}(m_{0u},m_{0d},m_{0s})$ is the current quark mass matrix and
\begin{eqnarray}
D_\mu=\partial_\mu+iQA_\mu,
\end{eqnarray}
is the covariant derivative with $A_\mu$ representing a PEM field and the charge matrix 
$$Q={\rm diag}(q_u,q_d,q_s)={\rm diag}({2\over 3},-{1\over 3},-{1\over 3})e.$$
Without loss of generality, we choose the gauge in Euclidean space as $A_\mu=(iEz,0,Bx,0)$. In the four-quark interaction terms with coupling $G$, $\lambda^0=\sqrt{2/3}~{\rm diag}(1,1,1)$ and $\lambda^i~(i=1,\dots,8)$ are Gell-Mann matrices in flavor space. Finally, in the 't Hooft term, the determinant is also performed in flavor space and the interaction vertices are $\Gamma^\pm=1\pm\gamma_5$ for right- and left-handed channels, respectively. 

In the setup $\mathbf{B\cdot E}\neq0$, we've already found $\pi^0$ and $\eta^0$ condensations  in two-flavor NJL models due to chiral anomaly~\cite{Cao:2015cka,Wang:2018gmj}. With the additional inclusion of strangeness, another neutral pseudoscalar
field, the $\eta^8$ meson, should also be expected to condensate in the same setup though the effect might be small. Thus, for a consistent exploration, we should set $\pi_{\rm f}^5=\langle\bar\psi_{\rm f}i\gamma^5\psi_{\rm f}\rangle/N_c$ to be nonzero besides the scalar condensates $\sigma_{\rm f}=\langle\bar\psi_{\rm f}\psi_{\rm f}\rangle/N_c$ with ${\rm f}={\rm u,d,s}$. To facilitate the study, we'd like firstly to reduce ${\cal L}_{\rm tH}$ to an effective form of four-fermion interactions in Hartree approximation. By following the derivations in Ref.~\cite{Klevansky:1992qe} where a quark-antiquark pair is contracted in each six-fermion interaction term, we immediately find
\begin{widetext}
\begin{eqnarray}
{\cal L}_{\rm tH}^4&=&-{K\over2}\sum_{s=\pm}\epsilon_{ijk}\epsilon_{imn}\langle\bar{\psi}^i\Gamma^s{\psi}^i\rangle(\bar{\psi}^j\Gamma^s{\psi}^m)(\bar{\psi}^k\Gamma^s{\psi}^n)\nonumber\\
&=&\!\!-N_c{K}\epsilon_{ijk}\epsilon_{imn}\left\{\sigma_i\left[(\bar{\psi}^j{\psi}^m)(\bar{\psi}^k{\psi}^n)
\!-\!(\bar{\psi}^ji\gamma^5{\psi}^m)(\bar{\psi}^ki\gamma^5{\psi}^n)\right]-2\pi_i^5(\bar{\psi}^ji\gamma^5{\psi}^m)(\bar{\psi}^k{\psi}^n)\right\}\nonumber\\
&=&-N_c{K\over6}\Big\{2\!\sum_{i=1}^3\sigma_{i}(\bar{\psi}\lambda^0\psi)^2\!-\!3\sigma_s\!\sum_{i=1}^3(\bar{\psi}\lambda^i\psi)^2\!-\!3\sigma_d\!\sum_{i=4}^5(\bar{\psi}\lambda^i\psi)^2
\!-\!3\sigma_u\!\sum_{i=6}^7(\bar{\psi}\lambda^i\psi)^2\!-\!\!\sum_{i=1}^3\! \sigma_i\Big[{(\bar{\psi}\lambda^8\psi)^2
\over \sqrt{3}/2\lambda^8_{ii}}\!+\!\sqrt{6}\lambda^8_{ii}(\bar{\psi}\lambda^0\psi)\nonumber\\
&&(\bar{\psi}\lambda^8\psi)+\sqrt{6}\lambda^3_{ii}(\bar{\psi}\lambda^3\psi)(\bar{\psi}\lambda^0\psi-\sqrt{2}\bar{\psi}\lambda^8\psi)\Big]-(\lambda^a\rightarrow i\lambda^a\gamma^5)\Big\}\nonumber\\
&&+N_c{K\over3}\Big\{2\sum_{i=1}^3\pi^5_i(\bar{\psi}\lambda^0\psi)(\bar{\psi}i\lambda^0\gamma^5\psi)
\!-\!3\pi^5_s\sum_{i=1}^3(\bar{\psi}\lambda^i\psi)(\bar{\psi}i\lambda^i\gamma^5\psi)
-3\pi^5_d\sum_{i=4}^5(\bar{\psi}\lambda^i\psi)(\bar{\psi}i\lambda^i\gamma^5\psi)-3\pi^5_u\sum_{i=6}^7
(\bar{\psi}\lambda^i\psi)\nonumber\\
&&(\bar{\psi}i\lambda^i\gamma^5\psi)\!-\!\sum_{i=1}^3\! \pi^5_i\Big[{(\bar{\psi}\lambda^8\psi)(\bar{\psi}i\lambda^8\gamma^5\psi)\over \sqrt{3}/2\lambda^8_{ii}}\!+\!{\sqrt{6}\over 2}\lambda^8_{ii}\Big((\bar{\psi}\lambda^0\psi)(\bar{\psi}i\lambda^8\gamma^5\psi)+(\bar{\psi}\lambda^8\psi)(\bar{\psi}i\lambda^0\gamma^5\psi)\Big)+{\sqrt{6}\over 2}\lambda^3_{ii}\Big((\bar{\psi}\lambda^3\psi)\nonumber\\
&&(\bar{\psi}i\lambda^0\gamma^5\psi-\sqrt{2}\bar{\psi}i\lambda^8\gamma^5\psi)+(\bar{\psi}i\lambda^3\gamma^5\psi)(\bar{\psi}\lambda^0\psi-\sqrt{2}\bar{\psi}\lambda^8\psi)\Big)\Big]\Big\},\label{LtH4}
\end{eqnarray}
where $\epsilon_{ijk}$ is the Levi-Civita symbol with $\epsilon_{123}=1$ and we've used the Einstein summation convention for the flavor indices $i,j,k,m,n$ in the first two steps. The correspondences between $1,2,3$ and $u,d,s$ should be understood for the subscripts here and throughout all the rest of the paper. Substitute ${\cal L}_{\rm tH}$ in Eq.\eqref{LNJL} by ${\cal L}_{\rm tH}^4$, the effective Lagrangian density with only four-fermion interactions is
\begin{eqnarray}
{\cal L}_{\rm NJL}^4\!\!&=&\!\!\bar\psi(i\slashed{D}-m_0)\psi+\sum_{a=0}^8\left\{G_{aa}^-(\bar\psi\lambda^a\psi)^2+G_{aa}^{+}(\bar\psi i\gamma_5\lambda^a\psi)^2+G_{aa}^5[(\bar\psi\lambda^a\psi)(\bar\psi i\lambda^a\gamma^5\psi)+(\bar\psi i\lambda^a\gamma^5\psi)(\bar\psi\lambda^a\psi)]\right\}\nonumber\\
&&+\!\sum_{a,b=0,3,8}^{a\neq b}\!\!\left\{G_{ab}^-(\bar\psi\lambda^a\psi)(\bar\psi\lambda^b\psi)+G_{ab}^+(\bar\psi i\lambda^a\gamma^5\psi)(\bar\psi i\lambda^b\gamma^5\psi)+G_{ab}^5[(\bar\psi \lambda^a\psi)(\bar\psi i\lambda^b\gamma^5\psi)+(\bar\psi i\lambda^a\gamma^5\psi)(\bar\psi \lambda^b\psi)]\right\},\label{LNJL4}
\end{eqnarray}
where the effective and symmetric couplings for the pure scalar-pseudoscalar channels $G_{ab}^\mp$ and the mixing ones $G_{ab}^5$ are respectively:
\begin{eqnarray}\label{Gelements}
&&G_{00}^\mp=G\mp N_c{K\over3}\sum_{i=1}^3\sigma_i,~G_{11}^\mp=G_{22}^\mp=G_{33}^\mp=G\pm N_c{K\over2}\sigma_s,~G_{44}^\mp=G_{55}^\mp=G\pm N_c{K\over2}\sigma_d,~G_{66}^\mp=G_{77}^\mp=G\pm N_c{K\over2}\sigma_u,\nonumber\\
&&G_{88}^\mp=G\mp N_c{K\over6}(\sigma_s-2\sigma_u-2\sigma_d),~G_{08}^\mp=\mp N_c{\sqrt{2}K\over12}(2\sigma_s-\sigma_u-\sigma_d),~G_{38}^\mp=-\sqrt{2}G_{03}^\mp=\mp N_c{\sqrt{3}K\over6}(\sigma_u-\sigma_d);\nonumber\\
&&G_{00}^5=N_c{K\over3}\sum_{i=1}^3\pi_i^5,~G_{11}^5=G_{22}^5=G_{33}^5=-N_c{K\over2}\pi_s^5,~G_{44}^5=G_{55}^5=-N_c{K\over2}\pi_d^5,~G_{66}^5=G_{77}^5=-N_c{K\over2}\pi_u^5,\nonumber\\
&&G_{88}^5=N_c{K\over6}(\pi^5_s-2\pi^5_u-2\pi^5_d),~G_{08}^5=N_c{\sqrt{2}K\over12}(2\pi^5_s-\pi^5_u-\pi^5_d),~G_{38}^5=-\sqrt{2}G_{03}^5=N_c{\sqrt{3}K\over6}(\pi^5_u-\pi^5_d).
\end{eqnarray}
\end{widetext}
Notice that pseudoscalar condensates further develop couplings between the scalar channels (mediated by $"\sigma,a_0,a_8"$ mesons) and pseudoscalar channels (mediated by $"\pi^0,\eta_0,\eta_8"$ mesons), thus these isospin-parity eigenstates will mix with each other in the mass eigenstates. 

Armed with the reduced Lagrangian density, all the necessary analytic derivations are just parallel to those given in two-flavor NJL models~\cite{Cao:2015cka,Wang:2018gmj}. By contracting a quark-antiquark pair further in each isospin diagonal interaction term of Eq.\eqref{LNJL4}, the effective mass and pion condensate of each quark flavor are respectively:
\begin{eqnarray}
m_i^*&=&m_{0i}-4N_cG\sigma_i+2N_c^2K(\sigma_j\sigma_k\!-\!\pi^5_j\pi^5_k),\nonumber\\
\pi_i^*&=&-4N_cG\pi_i^5-2N_c^2K(\sigma_j\pi^5_k\!+\!\pi^5_j\sigma_k)\label{mpi}
\end{eqnarray}
with $i\neq j\neq k$. The $G$ and $K$ dependent terms in Eq.\eqref{mpi} correspond to the $U_A(1)$ symmetric and anomalous violating interactions, respectively. Then, the gap equations are given through the following six self-consistent conditions~\cite{Klevansky:1992qe}:
\begin{eqnarray}
\sigma_i&=&\langle\bar{\psi}^i{\psi}^i\rangle/N_c=-{\rm tr} S_i,\label{sigmai}\\
\pi_i^5&=&\langle\bar{\psi}^ii\gamma^5{\psi}^i\rangle/N_c=-{\rm tr}~ i\gamma^5S_i\label{pii5}
\end{eqnarray}
where ${\cal S}_i(x)=-\left[i{\slashed D}_i-m_i^*-i\gamma^5\pi_i^*\right]^{-1}$ is the effective propagator of a given quark flavor in PEM field and the trace should be taken over the spinor and coordinate spaces. With only neutral condensates involved,get the propagators can be evaluated independently for quarks with different flavors or colors. In principle, we should work in in-in formalism when electric field is involved~\cite{Cao:2019hku,Copinger:2018ftr}, here we simply adopt the simpler in-out formalism to avoid numerical difficulties.

In Euclidean space, the effective quark propagator has already been evaluated in momentum space as~\cite{Miransky:2015ava}
\begin{widetext}
\begin{eqnarray}
\hat{\cal S}_{\rm f}({p})
&=&i\int_0^\infty {ds}\exp\big\{-i{M_{\rm f}^*}^2s-{i\tanh(q_{\rm f}\bar{I}_2s)\over q_{\rm f}\bar{I}_2}({p}_4^2+p_3^2)-{i\tan(q_{\rm f}\bar{I}_2s)\over q_{\rm f}\bar{I}_2}(p_1^2+p_2^2)\Big\}\big[m_{\rm f}^*-i\gamma^5\pi^*_{\rm f}\nonumber\\
&&-\gamma^4(p_4-{i\tanh(q_{\rm f}\bar{I}_2s)}p_3)-\gamma^3(p_3+{i\tanh(q_{\rm f}\bar{I}_2s)}p_4)-\gamma^2(p_2+{\tan(q_{\rm f}\bar{I}_2s)}p_1)-\gamma^1(p_1-{\tan(q_{\rm f}\bar{I}_2s)}p_2)\big]\nonumber\\
&&\Big[1+{i\gamma^5\tan(q_{\rm f}\bar{I}_2s)\tanh(q_{\rm f}\bar{I}_2s)}
+{\gamma^1\gamma^2\tan(q_{\rm f}\bar{I}_2s)}+{i\gamma^4\gamma^3\tanh(q_{\rm f}\bar{I}_2s)}\Big].
\end{eqnarray}
\end{widetext}
Here, we define the chiral mass $M_{\rm f}^*=({m_{\rm f}^*}^2+{\pi_{\rm f}^*}^2)^{1/2}$ and the field strength is chosen to be $E=B=\bar{I}_2$ without loss of generality. Inserting it into Eqs.\eqref{sigmai} and \eqref{pii5} and transforming the integral variable $s\rightarrow-is$, the explicit forms of the gap equations are:
\begin{eqnarray}
-\sigma_{\rm f}
&=&{m_{\rm f}^*\over4\pi^2}\int_0^\infty {ds\over s^2}{(q_{\rm f}\bar{I}_2s)^2e^{-{M_{\rm f}^*}^2s}
\over\tan(q_{\rm f}\bar{I}_2s)\tanh(q_{\rm f}\bar{I}_2s)}-{q_{\rm f}^2I_2\over4\pi^{2}}{{\pi_{\rm f}^*}\over{{M_{\rm f}^*}^2}}
,\nonumber\\\label{m0gap}\\
-\pi^5_{\rm f}
&=&{\pi_{\rm f}^*\over4\pi^2}\int_0^\infty {ds\over s^2}{(q_{\rm f}\bar{I}_2s)^2e^{-{M_{\rm f}^*}^2s}
\over\tan(q_{\rm f}\bar{I}_2s)\tanh(q_{\rm f}\bar{I}_2s)}+{q_{\rm f}^2I_2\over4\pi^{2}}{{m_{\rm f}^*}\over{{M_{\rm f}^*}^2}}
,\nonumber\\\label{pi0gap}
\end{eqnarray}
where $I_2={\bf E\cdot B}$ is the second Lorentz invariant. These forms are divergent, we would like to use the vacuum regularization scheme~\cite{Cao:2014uva} to proceed to numerical evaluations, then the gap equations become:
\begin{eqnarray}
-\sigma_{\rm f}
&=&{m_{\rm f}^*}F_\Lambda(M_{\rm f}^*)-{q_{\rm f}^2I_2\over4\pi^{2}}{{\pi_{\rm f}^*}\over{{M_{\rm f}^*}^2}},\label{mgap}\\
-\pi^5_{\rm f}&=&{\pi_{\rm f}^*}F_\Lambda(M_{\rm f}^*)+{q_{\rm f}^2I_2\over4\pi^{2}}{{m_{\rm f}^*}\over{{M_{\rm f}^*}^2}}\label{pigap}
\end{eqnarray}
with the auxiliary function defined as
\begin{widetext}
\begin{eqnarray}
F_\Lambda(M_{\rm f}^*)&=&{{M_{\rm f}^*}\over2\pi^2}\Big[\Lambda\Big({1\!+\!{\Lambda^2\over {M_{\rm f}^*}^2}}\Big)^{1\over2}\!-\!{M_{\rm f}^*}\ln\Big({\Lambda\over {M_{\rm f}^*}}
\!+\!\Big({1\!+\!{\Lambda^2\over {M_{\rm f}^*}^2}}\Big)^{1\over2}\Big)\Big]\!+\!{1\over4\pi^2}\int_0^\infty {ds\over s^2}e^{-{M_{\rm f}^*}^2s}\Big[{(q_{\rm f}\bar{I}_2s)^2
\over\tan(q_{\rm f}\bar{I}_2s)\tanh(q_{\rm f}\bar{I}_2s)}\!-\!1\Big].\nonumber
\end{eqnarray}
Getting rid of the cutoff terms, model-independent results follow Eqs.\eqref{mgap} and \eqref{pigap}:
 \begin{eqnarray}
\sigma_{\rm f}{\pi_{\rm f}^*}-\pi^5_{\rm f}{m_{\rm f}^*}={q_{\rm f}^2I_2\over4\pi^{2}}.\label{Min}
\end{eqnarray}
The two-flavor results in Ref.~\cite{Wang:2018gmj} can then be reproduced by setting $K=0$.

Finally, the overall thermodynamic potential can be derived consistently as
\begin{eqnarray}
\Omega&=&2N_cG\sum_{{\rm f}=u,d,s}(\sigma_{\rm f}^2+(\pi_{\rm f}^5)^2)\!-\!2N_c^2K(2\prod_{{\rm f}=u,d,s}\sigma_{\rm f}\!-\!\epsilon_{ijk}^2\sigma_i\pi_j^5\pi_k^5)\!-\!\sum_{{\rm f}=u,d,s}\left\{{{{M_{\rm f}^*}^3}\over8\pi^2}\Big[\Lambda\Big(1\!+\!{2\Lambda^2\over {{M_{\rm f}^*}^2}}\Big)\Big({1\!+\!{\Lambda^2\over {M_{\rm f}^*}^2}}\Big)^{1\over2}\!-\!\right.\nonumber\\
&&\left.{M_{\rm f}^*}\ln\Big({\Lambda\over {M_{\rm f}^*}}
\!+\!\Big({1\!+\!{\Lambda^2\over {M_{\rm f}^*}^2}}\Big)^{1\over2}\Big)\Big]\!+\!{1\over8\pi^2}\!\int_0^\infty \!{ds\over s^3}e^{-{M_{\rm f}^*}^2s}\Big[{(q_{\rm f}\bar{I}_2s)^2
\over\tan(q_{\rm f}\bar{I}_2s)\tanh(q_{\rm f}\bar{I}_2s)}\!-\!1\Big]\!-\!{q_{\rm f}^2I_2\over4\pi^{2}}\tan^{-1}\Big({{\pi_{\rm f}^*}\over {m_{\rm f}^*}}\Big)\right\}\label{Omega}
\end{eqnarray}
\end{widetext}
 by combining the integrations over ${m_{\rm f}^*}$ of Eq.(\ref{mgap}) and  the integrations over ${\pi_{\rm f}^*}$ of Eq.(\ref{pigap}). The expression $\tan^{-1}\Big({{\pi_{\rm f}^*}\over {m_{\rm f}^*}}\Big)$ in the last term of Eq.\eqref{Omega} are just chiral angles for different flavors~\cite{Wang:2017pje} and would change randomly beyond the ends of corresponding chiral rotations~\cite{Cao:2019hku}. Since the quark mass ${m_{\rm f}^*}$ is usually positive before the completion of chiral rotation,  this term indicates that pion condensate ${\pi_{\rm f}^*}$ always prefers the same sign as $I_2$ if QED anomaly dominates over the QCD one, regardless of the charge. This has already been well checked in the two-flavor $U_A(1)$ symmetric NJL model~\cite{Wang:2018gmj}. 

\subsection{Polarization functions}\label{Pfunction}

Now, we focus on the collective modes, especially the neutral scalars $\sigma,a_0,a_8$ and pseudoscalars $\pi^0,\eta_0,\eta_8$ which mix with each other. In the well-known random phase approximation~\cite{Klevansky:1992qe}, the kinetic terms of and mixings among these mesons are completely determined by the $6\times6$ polarization function matrix, which can be more conveniently derived from the reduced Lagrangian density Eq.\eqref{LNJL4}. Actually, only $21$ elements of the matrix are independent due to the transpose symmetry. To obtain these functions, we evaluate the traces over spinor space for each flavor first:
\begin{widetext}
\begin{eqnarray}
T_{\rm f}^-&\equiv&-{\rm tr}\hat{S}_{\rm f}(p+q/2)\hat{S}_{\rm f}(p-q/2)\nonumber\\
&=&4\int_0^\infty {ds}\int_0^\infty {ds'}\exp\left\{\!-\!i{M_{\rm f}^*}^2(s\!+\!s')\!-\!{i~{\rm th}\over q_{\rm f}\bar{I}_2}({{p}_4^+}^2\!+\!{p_3^+}^2)\!-\!{i~{\rm t}\over q_{\rm f}\bar{I}_2}({p_1^+}^2\!+\!{p_2^+}^2)\!-\!{i~{\rm th}'\over q_{\rm f}\bar{I}_2}({p_4^-}^2\!+\!{p_3^-}^2)\!-\!{i~{\rm t}'\over q_{\rm f}\bar{I}_2}({p_1^-}^2\!+\!{p_2^-}^2)\right\}\nonumber\\
&&\Big[({m_{\rm f}^*}^2\!-\!{\pi_{\rm f}^*}^2)(1\!-\!{\rm t}~{\rm t}')(1\!+\!{\rm th}~{\rm th}')\!+\!2{m_{\rm f}^*}{\pi_{\rm f}^*}({\rm t}\!+\!{\rm t}')({\rm th}\!+\!{\rm th}')\!-\!({{P}_4^+}{{P}_4^-}\!+\!{{P}_3^+}{{P}_3^-})(1\!-\!{\rm t}~{\rm t}')(1\!-\!{\rm th}~{\rm th}')\!+\!i({{P}_4^+}{{P}_3^-}\!-\!\nonumber\\
&&{{P}_3^+}{{P}_4^-})(1\!-\!{\rm t}~{\rm t}')({\rm th}\!-\!{\rm th}')\!-\!({{P}_2^+}{{P}_2^-}\!+\!{{P}_1^+}{{P}_1^-})(1\!+\!{\rm t}~{\rm t}')(1\!+\!{\rm th}~{\rm th}')\!+\!({{P}_1^+}{{P}_2^-}\!-\!{{P}_2^+}{{P}_1^-})({\rm t}\!-\!{\rm t}')(1\!+\!{\rm th}~{\rm th}')\Big],\\
T_{\rm f}^+&\equiv&-{\rm tr}[i\gamma^5\hat{S}_{\rm f}(p+q/2)i\gamma^5\hat{S}_{\rm f}(p-q/2)]\nonumber\\
&=&4\int_0^\infty {ds}\int_0^\infty {ds'}\exp\left\{\!-\!i{M_{\rm f}^*}^2(s\!+\!s')\!-\!{i~{\rm th}\over q_{\rm f}\bar{I}_2}({{p}_4^+}^2\!+\!{p_3^+}^2)\!-\!{i~{\rm t}\over q_{\rm f}\bar{I}_2}({p_1^+}^2\!+\!{p_2^+}^2)\!-\!{i~{\rm th}'\over q_{\rm f}\bar{I}_2}({p_4^-}^2\!+\!{p_3^-}^2)\!-\!{i~{\rm t}'\over q_{\rm f}\bar{I}_2}({p_1^-}^2\!+\!{p_2^-}^2)\right\}\nonumber\\
&&\Big[-({m_{\rm f}^*}^2\!-\!{\pi_{\rm f}^*}^2)(1\!-\!{\rm t}~{\rm t}')(1\!+\!{\rm th}~{\rm th}')\!-\!2{m_{\rm f}^*}{\pi_{\rm f}^*}({\rm t}\!+\!{\rm t}')({\rm th}\!+\!{\rm th}')\!-\!({{P}_4^+}{{P}_4^-}\!+\!{{P}_3^+}{{P}_3^-})(1\!-\!{\rm t}~{\rm t}')(1\!-\!{\rm th}~{\rm th}')\!+\!i({{P}_4^+}{{P}_3^-}\!-\!\nonumber\\
&&{{P}_3^+}{{P}_4^-})(1\!-\!{\rm t}~{\rm t}')({\rm th}\!-\!{\rm th}')\!-\!({{P}_2^+}{{P}_2^-}\!+\!{{P}_1^+}{{P}_1^-})(1\!+\!{\rm t}~{\rm t}')(1\!+\!{\rm th}~{\rm th}')\!+\!({{P}_1^+}{{P}_2^-}\!-\!{{P}_2^+}{{P}_1^-})({\rm t}\!-\!{\rm t}')(1\!+\!{\rm th}~{\rm th}')\Big],\\
T_{\rm f}^5&\equiv&-{\rm tr}[\hat{S}_{\rm f}(p+q/2)i\gamma^5\hat{S}_{\rm f}(p-q/2)]=-{\rm tr}[i\gamma^5\hat{S}_{\rm f}(p-q/2)\hat{S}_{\rm f}(p+q/2)]\nonumber\\
&=&4\int_0^\infty {ds}\int_0^\infty {ds'}\exp\left\{\!-\!i{M_{\rm f}^*}^2(s\!+\!s')\!-\!{i~{\rm th}\over q_{\rm f}\bar{I}_2}({{p}_4^+}^2\!+\!{p_3^+}^2)\!-\!{i~{\rm t}\over q_{\rm f}\bar{I}_2}({p_1^+}^2\!+\!{p_2^+}^2)\!-\!{i~{\rm th}'\over q_{\rm f}\bar{I}_2}({p_4^-}^2\!+\!{p_3^-}^2)\!-\!{i~{\rm t}'\over q_{\rm f}\bar{I}_2}({p_1^-}^2\!+\!{p_2^-}^2)\right\}\nonumber\\
&&\Big[2{m_{\rm f}^*}{\pi_{\rm f}^*}(1\!-\!{\rm t}~{\rm t}')(1\!+\!{\rm th}~{\rm th}')\!-\!({m_{\rm f}^*}^2\!-\!{\pi_{\rm f}^*}^2)({\rm t}\!+\!{\rm t}')({\rm th}\!+\!{\rm th}')\!+\!({{P}_4^+}{{P}_4^-}\!+\!{{P}_3^+}{{P}_3^-})({\rm t}\!+\!{\rm t}')({\rm th}\!-\!{\rm th}')\!-\!i({{P}_4^+}{{P}_3^-}\!-\!{{P}_3^+}{{P}_4^-})\nonumber\\
&&({\rm t}+{\rm t}')(1-{\rm th}~{\rm th}')+({{P}_2^+}{{P}_2^-}+{{P}_1^+}{{P}_1^-})({\rm t}-{\rm t}')({\rm th}+{\rm th}')+({{P}_1^+}{{P}_2^-}-{{P}_2^+}{{P}_1^-})(1+{\rm t}~{\rm t}')({\rm th}+{\rm th}')\Big],
\end{eqnarray}
where the following denotations should be understood: 
\begin{eqnarray}
&&{\rm t}=\tan(q_{\rm f}\bar{I}_2s),{\rm th}=\tanh(q_{\rm f}\bar{I}_2s),{\rm t}'=\tan(q_{\rm f}\bar{I}_2s'),{\rm th}'=\tanh(q_{\rm f}\bar{I}_2s'), p_\mu^\pm=p_\mu\pm q_\mu/2,\nonumber\\
&& P_4^+=p_4^+-ip_3^+{\rm th},P_3^+=p_3^++ip_4^+{\rm th},P_2^+=p_2^++p_1^+{\rm t},P_1^+=p_1^+-p_2^+{\rm t}, \nonumber\\ 
&&P_4^-=p_4^--ip_3^-{\rm th}',P_3^-=p_3^-+ip_4^-{\rm th}',P_2^-=p_2^-+p_1^-{\rm t}', P_1^-=p_1^--p_2^-{\rm t}'.\nonumber
\end{eqnarray} 

For the evaluations of pole masses, we set the three-momentum  ${\bf q=0}$, then the four-momentum integrated forms of the trace functions $\Pi_{\rm f}^n(q_4)\equiv-\int{d^4p\over(2\pi)^4}T_{\rm f}^n(p,q_4)~(n=\pm,5)$ are respectively:
\begin{eqnarray}
\Pi_{\rm f}^\mp(q_4)
&=&{1\over4\pi^2}\!\int_0^\infty\! {ds}\!\int_0^\infty\! {ds'}{q_{\rm f}\bar{I}_2\over{\rm th}\!+\!{\rm th}'}{q_{\rm f}\bar{I}_2\over{\rm t}\!+\!{\rm t}'}~e^{-i{M_{\rm f}^*}^2(s+s')+{i({\rm th}-{\rm th}')^2\over q_{\rm f}\bar{I}_2({\rm th}+{\rm th}')}{q_4^2\over4}}~\Big[\!\pm\!({m_{\rm f}^*}^2\!-\!{\pi_{\rm f}^*}^2)(1\!-\!{\rm t}~{\rm t}')(1\!+\!{\rm th}~{\rm th}')\!\pm\!2{m_{\rm f}^*}{\pi_{\rm f}^*}({\rm t}\!+\!{\rm t}')\nonumber\\
&&\ \ \  ({\rm th}\!+\!{\rm th}')\!-\!\Big({q_{\rm f}\bar{I}_2\over i({\rm th}\!+\!{\rm th}')}\!-\!{{\rm th}~{\rm th}'\over({\rm th}\!+\!{\rm th}')^2}{q_4^2}\Big)(1\!-\!{\rm t}~{\rm t}')(1\!-\!{\rm th}^2)(1\!-\!{{\rm th}'}^2)\!-\!{q_{\rm f}\bar{I}_2\over i({\rm t}\!+\!{\rm t}')}(1\!+\!{\rm th}~{\rm th}')(1\!+\!{\rm t}^2)(1\!+\!{{\rm t}'}^2)\Big],\\
\Pi_{\rm f}^5(q_4)
&=&{1\over4\pi^2}\!\int_0^\infty\! {ds}\!\int_0^\infty\! {ds'}{q_{\rm f}\bar{I}_2\over{\rm th}\!+\!{\rm th}'}{q_{\rm f}\bar{I}_2\over{\rm t}\!+\!{\rm t}'}~e^{-i{M_{\rm f}^*}^2(s+s')+{i({\rm th}-{\rm th}')^2\over q_{\rm f}\bar{I}_2({\rm th}+{\rm th}')}{q_4^2\over4}}~\Big[2{m_{\rm f}^*}{\pi_{\rm f}^*}(1\!-\!{\rm t}~{\rm t}')(1\!+\!{\rm th}~{\rm th}')\!-\!({m_{\rm f}^*}^2\!-\!{\pi_{\rm f}^*}^2)({\rm t}\!+\!{\rm t}')\nonumber\\
&&\ \ \  ({\rm th}\!+\!{\rm th}')\Big].
\end{eqnarray}
Transform the integral variables as $s\rightarrow s(1+u)/2, s'\rightarrow s(1-u)/2$  and regularize the functions $\Pi_{\rm f}^n$ with the help of the corresponding terms in the limit $\bar{I}_2\rightarrow 0$, we find~\cite{Cao:2019hku}
\begin{eqnarray}
\Pi_{\rm \Lambda f}^\mp(q_4)
&=&{q_{\rm f}^2I_2\over8\pi^2}\!\int_0^\infty\!\! s{ds}\!\int_{-1}^1\! {du}~e^{-i{M_{\rm f}^*}^2s+{i(\bar{{\rm th}}_{\rm f}-\bar{{\rm th}}_{\rm f}')^2\over q_{\rm f}\bar{I}_2(\bar{{\rm th}}_{\rm f}+\bar{{\rm th}}_{\rm f}')}{q_4^2\over4}}~\Bigg[{\pm({m_{\rm f}^*}^2-{\pi_{\rm f}^*}^2)\over\tan(q_{\rm f}\bar{I}_2s)\tanh(q_{\rm f}\bar{I}_2s)}\!\pm\!2{m_{\rm f}^*}{\pi_{\rm f}^*}\!-\!{q_{\rm f}\bar{I}_2\left(1-{i\,\bar{{\rm th}}_{\rm f}~\bar{{\rm th}}_{\rm f}'\over q_{\rm f}\bar{I}_2(\bar{{\rm th}}_{\rm f}+\bar{{\rm th}}_{\rm f}')}{q_4^2}\right)\over i\,\tan(q_{\rm f}\bar{I}_2s)\sinh^2(q_{\rm f}\bar{I}_2s)}\nonumber\\
&&-{q_{\rm f}\bar{I}_2\over i\,\tanh(q_{\rm f}\bar{I}_2s)}{1\over\sin^2(q_{\rm f}\bar{I}_2s)}\Bigg]-(\bar{I}_2\rightarrow0)-8\int^\Lambda{d^3p\over(2\pi)^3}{E_{\rm f}^2(p)-{1\over2}\left[{M_{\rm f}^*}^2\pm({m_{\rm f}^*}^2-{\pi_{\rm f}^*}^2)\right]\over E_{\rm f}(p)(q_4^2+4E_{\rm f}^2(p))},\\
\Pi_{\rm  \Lambda f}^5(q_4)
&=&{q_{\rm f}^2I_2\over8\pi^2}\!\int_0^\infty\!\! s{ds}\!\int_{-1}^1\! {du}~e^{-i{M_{\rm f}^*}^2s+{i(\bar{{\rm th}}_{\rm f}-\bar{{\rm th}}_{\rm f}')^2\over q_{\rm f}\bar{I}_2(\bar{{\rm th}}_{\rm f}+\bar{{\rm th}}_{\rm f}')}{q_4^2\over4}}~\left[{2{m_{\rm f}^*}{\pi_{\rm f}^*}\over\tan(q_{\rm f}\bar{I}_2s)\tanh(q_{\rm f}\bar{I}_2s)}-({m_{\rm f}^*}^2-{\pi_{\rm f}^*}^2)\right]-(\bar{I}_2\rightarrow0)\nonumber\\
&&+8\int^\Lambda{d^3p\over(2\pi)^3}{{m_{\rm f}^*}{\pi_{\rm f}^*}\over E_{\rm f}(p)(q_4^2+4E_{\rm f}^2(p))},
\end{eqnarray}
where the dispersion relationship in the cutoff terms is $E_{\rm f}(p)=(p^2+{M_{\rm f}^*}^2)^{1/2}$ and we define $\bar{{\rm th}}_{\rm f}=\tanh((1+u)q_{\rm f}\bar{I}_2s/2)$ and $\bar{{\rm th}}_{\rm f}'=\tanh((1-u)q_{\rm f}\bar{I}_2s/2)$ for brevity. Note that the integrations over the Euclidean energy $p_4$ have already been carried out in the vacuum terms. 
\end{widetext}

For convenience, we set three diagonal matrices 
$$\Pi^n(q_4)={\rm diag}(\Pi_{\rm \Lambda u}^n(q_4),\Pi_{\rm \Lambda d}^n(q_4),\Pi_{\rm \Lambda s}^n(q_4))~(n=\pm,5),$$
then the polarization functions among the isospin eigenstates, that is,
 \begin{eqnarray}
 \Pi^{-}_{ij}\!&\equiv&\!\int{d^4p\over(2\pi)^4}{\rm Tr}\hat{S}(p+q/2)\lambda^i\hat{S}(p-q/2)\lambda^j,\\
 \Pi^{+}_{ij}\!&\equiv&\!\int{d^4p\over(2\pi)^4}{\rm Tr}\hat{S}(p+q/2)i\gamma^5\lambda^i\hat{S}(p-q/2)i\gamma^5\lambda^j,\\
 \Pi^{5}_{ij}\!&\equiv&\!\int{d^4p\over(2\pi)^4}{\rm Tr}\hat{S}(p+q/2)\lambda^i\hat{S}(p-q/2)i\gamma^5\lambda^j,
 \end{eqnarray}
  can be evaluated directly through traces over the flavor space:
\begin{eqnarray}
\Pi^{n}_{ij}(q_4)={N_c}{\rm tr_f}[\lambda^i\Pi^n(q_4)\lambda^j].
\end{eqnarray}
Since $\Pi^n$ and $\lambda^i~(i=0,3,8)$ are all diagonal, $\Pi^{n}_{ij}$ are all symmetric under the exchange of the subscript indices $i$ and $j$. Gather all the polarization functions into a matrix for the whole scalar-pseudoscalar sector with the generalized meson field $(\sigma,a_0,a_8,\eta_0,\pi^0,\eta_8)^T$, we have 
\begin{eqnarray}
{\Pi}_{\rm SP}(q_4)=\left(\begin{array}{cccccc}
{\Pi}_{00}^-&\Pi^{-}_{03}&\Pi^{-}_{08}&\Pi^{5}_{00}&\Pi^{5}_{03}&\Pi^{5}_{08}\\
{\Pi}_{03}^-&{\Pi}^{-}_{33}&\Pi^{-}_{38}&\Pi^{5}_{03}&\Pi^{5}_{33}&\Pi^{5}_{38}\\
{\Pi}_{08}^-&\Pi^{-}_{38}&{\Pi}^{-}_{88}&\Pi^{5}_{08}&\Pi^{5}_{38}&\Pi^{5}_{88}\\
{\Pi}_{00}^5&\Pi^{5}_{03}&\Pi^{5}_{08}&{\Pi}^{+}_{00}&\Pi^{+}_{03}&\Pi^{+}_{08}\\
{\Pi}_{03}^5&\Pi^{5}_{33}&\Pi^{5}_{38}&\Pi^{+}_{03}&{\Pi}^{+}_{33}&\Pi^{+}_{38}\\
{\Pi}_{08}^5&\Pi^{5}_{38}&\Pi^{5}_{88}&\Pi^{+}_{08}&\Pi^+_{38}&{\Pi}^{+}_{88}
\end{array}\right).
\end{eqnarray}
Then, the corresponding propagator matrix is given by~\cite{Klimt:1989pm}
\bea
P_{\rm SP}(q_4)=[1+2{\cal G}_{\rm SP}{\Pi}_{\rm SP}(q_4)]^{-1}{\cal G}_{\rm SP},
\eea
where ${\cal G}_{\rm SP}$ is the coupling matrix with the elements $G^{n}_{ij}$ list in Eq.\eqref{Gelements} and arranged in the same order as $\Pi^{n}_{ij}$ in ${\Pi}_{\rm SP}(q_4)$. 
Eventually, the pole masses of the collective modes can be obtained by solving the equation: 
$$\det \left[1+2{\cal G}_{\rm SP}{\Pi}_{\rm SP}(iM_m)\right]=0,$$
and there are six independent solutions in principle.

\section{numerical results}\label{num}

\begin{figure}[!htb]
	\centering
	\includegraphics[width=0.45\textwidth]{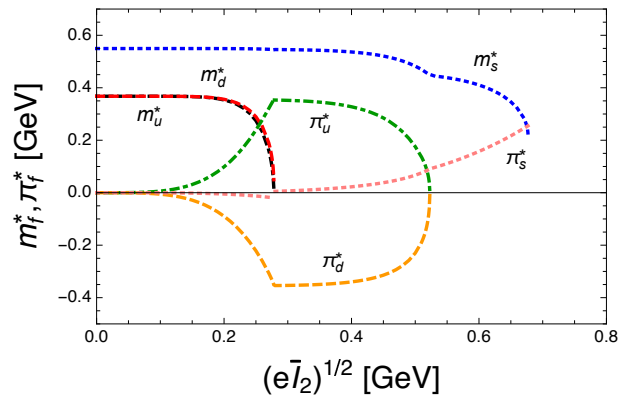}
	\caption{The effective masses $m^*_{\rm f}$ and pion condensates $\pi^*_{\rm f}$ as functions of the electromagnetic field $e\bar{I}_2$ for $u$ (dash-doted), $d$ (dashed) and $s$ (dotted) quarks, respectively. For brevity, the restorations of the corresponding order parameters to their current values are not shown beyond the critical points.}\label{condensates}	
\end{figure}

For the numerical calculations, we choose the following model parameters: $m_{0u}=m_{0d}=5.5~{\rm MeV}, m_{0s}=140.7~{\rm MeV}, \Lambda=602.3~{\rm MeV}, G\Lambda^2=1.835$ and $K\Lambda^5=12.36$~\cite{Rehberg:1995kh}. The effective masses and pion condensates of different flavors are solved self-consistently from Eqs.\eqref{mgap} and \eqref{pigap} and illuminated together in Fig.\ref{condensates}. The features of the more interesting pion condensates can be explained by adopting the cutoff-independent results Eq.\eqref{Min} and the fact that: in Eq.\eqref{mpi}, the $U_A(1)$ symmetric term of $\pi^*_{\rm f}$ dominates over the anomalous violating term as $G\gg |K\sigma_{\rm f}|$. Substitute Eq.\eqref{mpi} into Eq.\eqref{Min}, we get
\begin{eqnarray}
-4N_cG\pi^5_{\rm i}=4N_cG{{q_{\rm f}^2I_2\over4\pi^{2}}+2N_c^2K(\sigma_j\pi^5_k\!+\!\pi^5_j\sigma_k)\sigma_{\rm i}\over m_{0i}+2N_c^2K(\sigma_j\sigma_k\!-\!\pi^5_j\pi^5_k)}.\label{Mina}
\end{eqnarray}
As the denominator on the right-hand side of Eq.\eqref{Mina} is always positive, the sign of ${\pi_{\rm f}^*}$ just follows the numerator. For small EM field, the QCD anomaly dominates over the QED one, that is, the numerator is mainly determined by the second term.  According to the numerical results, $\sigma_{\rm i}$ are always negative and the QCD anomaly terms in the numerators are dominated by
\begin{eqnarray}
\sigma_s\pi^5_d, \sigma_s\pi^5_u, (\sigma_u\pi^5_d+\sigma_d\pi^5_u)
\end{eqnarray}
for $u,d,s$ quarks, respectively. Then, it is easy to find: the signs of ${\pi_{\rm u}^*}$ and  ${\pi_{\rm d}^*}$ are opposite to each other, and ${\pi_{\rm s}^*}<0$ as the numerical calculations show $\sigma_u\sim\sigma_d$ and $0<\pi^5_d<-\pi^5_u$. These are just the characters up to the end of first chiral rotation with the critical strength $e\bar{I}_2^{c_1}=(0.278~{\rm GeV})^2$. However, beyond this point with both $\sigma_u$ and $\sigma_d$ vanishing, $\pi^*_{\rm s}$ changes sign because the QCD anomaly loses significance for the strange quark. In one flavor case, we've explained that $\pi_{\rm u}$ condensate only stands for the magnitude of chiral condensate and the chiral angle can be randomly chosen~\cite{Cao:2019hku}. But for the three-flavor case, the constancy of $s$ quark dynamics requires $\sigma_u\sigma_d-\pi^5_u\pi^5_d$ to be almost a constant to keep ${m_{\rm s}^*}$ changing little. In other words, QCD anomaly correlates the chiral angles of $u$ and $d$ quarks in the three-flavor NJL model. Eventually, the chiral symmetry for $u$ and $d$ quarks restores at the second critical EM field $e\bar{I}_2^{c_2}=(0.523~{\rm GeV})^2$ and that for $s$ quark restores at the third critical field $e\bar{I}_2^{c_3}=(0.677~{\rm GeV})^2$. Note that the transitions at $e\bar{I}_2^{c_1}, e\bar{I}_2^{c_2}$ and $e\bar{I}_2^{c_3}$ are weak first order, second order and strong first order, respectively. Due to the large current mass of $s$ quark, the corresponding end of chiral rotation and chiral restoration point mergers into one, that is, $e\bar{I}_2^{c_3}$.

There is one thing needs to be clarified in external EM field: the flavor separations of the collective modes, which were usually assumed in LQCD simulations at large magnetic field~\cite{Hidaka:2012mz,Bali:2017ian,Ding:2020jui}. We rearrange the interaction terms of Eq.\eqref{LNJL4} in flavor eigen-channels and find the couplings of mixing terms to be:
$$G^{\mp}_{ij}=\mp K N_c \sigma_k,\ G^{5}_{ij}=K N_c \pi_k$$
with $i\neq j\neq k$. So, $u$ and $d$ quarks never separate from each other up to $e\bar{I}_2^{c_3}$ beyond which all the condensates almost vanish; and neither does $s$ quark from them up to $e\bar{I}_2^{c_2}$ where chiral symmetry for $u$ and $d$ quarks is eventually restored. In real QCD, the effective couplings would decrease with the EM field due to asymptotic freedom and the flavor separations might occur earlier.

\begin{figure}[!htb]
	\centering
	\includegraphics[width=0.42\textwidth]{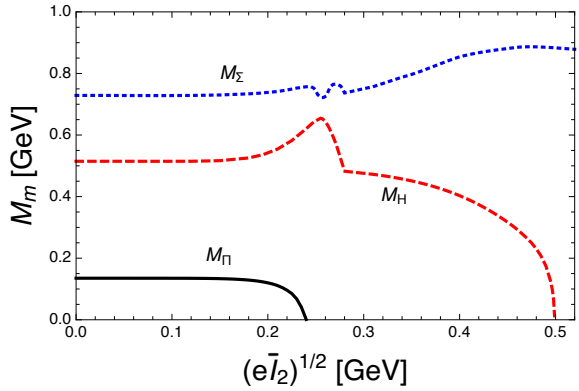}
	\caption{The meson masses as functions of the electromagnetic field $e\bar{I}_2$ for the lowest-lying neutral collective modes $\Sigma, \Pi$ and $H$ in the scalar-pseudoscalar sector.}\label{mesons}
\end{figure}

The masses of the lowest-lying collective modes $\Sigma, \Pi$ and $H$ are shown in Fig.\ref{mesons}. Note that these modes separately correspond to $\sigma,\pi^0$ and $\eta$ mesons in the vanishing EM field limit. As we can see, the masses of $\Pi$ and $H$ modes reduce to zero around $e\bar{I}_2^{c_1}$ and $e\bar{I}_2^{c_2}$, which actually signal the instabilities induced by QCD and QED anomalies for $u,d$ quarks, respectively. There is an extra feature: Around $e\bar{I}_2^{c_1}$, the scalar-pseudoscalar components exchange between the mass eigenstates $\Sigma$ and $H$, which demonstrates itself through the peak-to-dip crossing structure.   Concretely, the component $\bar{u}i\gamma^5u+\bar{d}i\gamma^5d$ dominates in $\Sigma$ mode whereas $\bar{u}u+\bar{d}d$ dominates in $H$ mode beyond $e\bar{I}_2^{c_1}$. Furthermore, the peak right after the dips in $M_\Sigma$ is just the hierarchy of that found in the one-flavor case around $e\bar{I}_2^{c_1}$~\cite{Cao:2019hku}, and there is a second but soft one around $e\bar{I}_2^{c_2}$. 

\section{Conclusions}\label{conclusion}
The neutral chiral condensates and lowest-lying collective modes are explored in the three-flavor NJL model, which is characterized by the 't Hooft determinant for the QCD $U_A(1)$ anomaly. Three critical EM fields are found for the chiral condensates: $e\bar{I}_2^{c_1}$ is the end of chiral rotation of $u$ and $d$ quarks, $e\bar{I}_2^{c_2}$ is the chiral restoration point of $u$ and $d$ quarks, and $e\bar{I}_2^{c_3}$ is the chiral restoration point of $s$ quark. We analyze in great detail mathematically and find the signs of pion condensates $\pi^*_{\rm f}$ reflect the relative significances of QCD and QED anomalies: the opposite signs of $\pi^*_{\rm u}$ and $\pi^*_{\rm d}$ and the negativeness of $\pi^*_{\rm s}$ both indicate the dominance of QCD anomaly in the small EM field region, and the positiveness of $\pi^*_{\rm s}$ implies the dominance of QED anomaly for $s$ quark. Along with the changes of chiral condensates, the masses of the two lowest-lying modes $\Pi$ and $H$ reduce to zero at $e\bar{I}_2^{c_1}$ and $e\bar{I}_2^{c_2}$, which actually signal the instabilities induced by QCD and QED anomalies, respectively. Besides, there is a components-exchange structure for the two modes $\Sigma$ and $H$ with their masses close to each other. Since there is only QED anomaly in our previous work~\cite{Cao:2019hku},  different feature of the lightest meson near $e\bar{I}_2^{c_1}$ here can be considered as the signal of QCD anomaly dominance, that is, the absence of peak structure in $M_\Pi$. 

Finally, it should be pointed out that: As we've demonstrated that the in-in and in-out results are almost the same when the EM field is relatively small~\cite{Cao:2019hku}, the most important discoveries of this work are reliable around the first critical point $e\bar{I}_2^{c_1}$.

\emph{Acknowledgments}---
G.C. is supported by the National Natural Science Foundation of China with Grant No. 11805290 and Young Teachers Training Program of Sun Yat-sen University with Grant No. 19lgpy282.

\end{document}